\documentclass[10pt,aps,pre,twocolumn,superscriptaddress]{revtex4-1}
\usepackage{bm}
\usepackage{graphicx}
\usepackage{epsfig}
\usepackage{amsmath,amssymb}
\usepackage{color}
\usepackage{natbib}

\begin{document}
\title{ Delayed feedback control of active particles: a controlled journey towards the destination}

\author{S. M. J. Khadem*}
\affiliation{Institute of Physics, Technical University of Berlin,  Hardenbergstr. 36, D-10623 Berlin, Germany}
\author{Sabine H. L. Klapp}
\affiliation{Institute of Physics, Technical University of Berlin,  Hardenbergstr. 36, D-10623 Berlin, Germany}

\begin{abstract}
We explore theoretically the navigation of an active particle based on delayed
feedback control. The delayed feedback enters in our expression for the particle orientation which, for an active particle, determines (up to noise) the direction of motion in the next time step. Here we 
estimate the orientation by comparing the delayed position of the particle with the actual one. This method does not require  any real-time monitoring of the 
particle orientation and may thus be relevant also for controlling sub-micron sized particles, where the imaging process is not easily feasible. 
We apply the delayed feedback strategy to two experimentally relevant situations, namely, optical trapping and photon nudging. To investigate the performance of our strategy, we calculate the mean arrival time
analytically (exploiting a small-delay approximation) and by simulations.
\end{abstract}

  \maketitle
  
  {jebreiilkhadem@itp.tu-berlin.de}

\section{Introduction}
Navigating the motion of active (self-propelled) particles, which are capable of converting energy from their surrounding environment into
directed motion, is a task 
receiving increasing interest. From an applicational point of view, this problem is relevant, e.g., for targeted active drug delivery \cite{wan} 
and in the context
of robotics \cite{robot}, but also for the controlled assembly of microscale systems based on active agents \cite{kim}. More fundamentally, the 
question is
how to manipulate the motion of an autonomous object subject to random fluctuations, either for an individual agent or for an ensemble. Indeed, 
even without 
explicit external control, active particles show a very rich and intriguing collective behavior (see, e.g., \cite{mike,ramin}), including 
formation of spatio-temporal
patterns on the mesoscale 
\cite{heidi1,heidi2}. 

Depending on the type of active particle, different mechanisms of (single-particle) control have been proposed involving, e.g., magnetic fields \cite{tierno}, topographical
fields (e.g., channels or patterned walls), chemical gradients, and optical
and thermophoretic forces (for a review, see \cite{reva}). Some of these strategies are based on {\em feedback} mechanisms, where the propulsion velocity \cite{hauf} or 
the direction of motion of the particle are continuously adapted depending on its current
position and/or orientation. Indeed, feedback control \cite{f1} is a concept currently gaining growing attention  in various areas of colloidal transport, including 
transport of passive colloids \cite{fu,f9,f10,f11},
(thermophoretic) control of DNA molecules \cite{brauni}, manipulation for biomedical engineering \cite{f8},
and  control of active particles  \cite{f7, chicos2} . 
An example of feedback control in the area of active colloidal particles is the photon nudging method \cite{chicos2}, 
where the propulsion generated by a weak laser is turned on or off
when the swimmer moves towards or away from the target (for a theoretical description, see \cite{selm}). In the present
paper, we propose yet another method of feedback control which, as a crucial
new feature, involves a {\em time delay}.

Our motivation to do so is as follows. A key ingredient in
various of the above-mentioned feedback control strategies (such as, e.g., photon nudging) is the particle orientation as function of time, which determines the direction of motion. Monitoring the particle orientation can, however, be a difficult task, especially
for {\em small} (sub-micronsize)
particles. The purpose of the present paper is to explore, on a theoretical level, a feedback control based on an {\em approximate} 
orientation vector defined as the
difference between the actual particle position at time t, $\textbf{r}(t)$, and its position at an earlier
time, $\textbf{r}(t-\delta)$. Here, $\delta$ is the ''delay'' time. By this approximation,
only the position needs to be monitored (which makes the method also applicable to
active particles without intrinsic orientation such as active micro-emulsion droplets \cite{ref2}).
The approximate orientation vector is then used to predict the next step of the translational motion
of the particle. Clearly, this approximation requires the delay time to be smaller than the rotational
relaxation time, which is one of the key assumptions of our approach. 
However, given a sufficiently fast imaging device, this assumption should be not too restrictive. 
The other key assumption
is that we can {\em adapt} the particle's motility (speed of propulsion) by the intensity of a laser, similar
to what has been done in various experimental studies \cite{chicos2, chicos1}.
We model our control concept mathematically based on suitable Langevin equations, and we discuss, on a theoretical level, 
two examples of application.

 The first example is inspired by the optical trapping method \cite{ashk}, where a particle's position is manipulated
by a co-moving laser beam. For real active particles sensitive to light, the laser intensity influences not only the stiffness of the trap, but also the motility \cite{jiang,jia}.
Within the strategy proposed here, the motility is adapted (through the laser intensity) in response to the approximate orientation vector, i.e., the displacement 
$\textbf{r}(t)-\textbf{r}(t-\delta)$. This introduces a symmetry-breaking of the conventional isotropic translational diffusion. To judge the performance of this strategy we analyze the resulting mean arrival time, which the particle needs to travel from a starting point to a predefined target. 
To this end we employ both, numerical simulations of the full (delayed) equations of motion and analytical results obtained from a
coarse-grained theory.

As a second example of application, we consider theoretically a variant of the photon-nudging method, where the current orientation vector is again estimated
on the basis of the displacement 
$\textbf{r}(t)-\textbf{r}(t-\delta)$. 
We then adapt the motility to perform the navigation.

It should be noted that, due to the usage of the position at an earlier time, our control method falls into the class of {\em delayed feedback control} strategies, which are well established in the area of chaos control \cite{schol,atax}, e.g. in laser 
systems \cite{marini,schemi} and in chemical reaction networks \cite{yui,pogo,thanh}. On a theoretical level, time delay considerably complicates the mathematical treatment since the underlying stochastic equations become
non-Markovian in character. This leads, e.g., to an non-trivial (hierarchical) Fokker-Planck equation already in the single-particle case \cite{loosi}.
Here we circumvent this problem by assuming the delay to be small, allowing for an effectively Markovian treatment similar to that employed
in an earlier study on the use of (sensorial) delay for autonomous agents \cite{Mijalkov,leyman2018tuning}.

The remainder of this work is structured as follows: In section~\ref{s1}, we discuss those types of active particles for which 
our method seems applicable, and we briefly introduce active Brownian particles as a suitable mathematical
model. In section~\ref{s2}, 
we propose a delayed feedback strategy within the optical trapping method and analyze the proposed equations of motion by theory and numerical simulation. 
For our analytical treatment, we focus on small delay times and introduce
coarse-grained equations of motion focusing solely on the translational dynamics, following closely an approach suggested in Ref.~\cite{Mijalkov}. Based on this theory we then calculate 
the mean arrival time, which turns out to agree very well with corresponding results 
from numerical simulations. In section~\ref{nudging},  we combine the concepts of delayed feedback method and the photon nudging method, and investigate its applicability numerically. We also 
compare our results with corresponding numerical results from ''standard" photon nudging based on the true particle orientation.
Finally, a summary and outlook is given in section~\ref{s5}.
\section{Systems of interest and model}
\label{s1}
 In contrast to passive Brownian particles, active particles display a directed motion on timescales which are much longer than the diffusion time of a passive particle 
of the same size. 
Such a motion may be generated, for instance,  through a chemical reaction of the particle  with the surrounding environment (examples being gold-platinum and 
gold-nickel microrods in hydrogen proxide solutions \cite{pax,four}) 
or by an external field. Examples of the latter kind are chiral magnetic objects driven by a magnetic field \cite{ghosh} or metallic microrods driven by an acoustic  field \cite{wang1}.

Regardless of the origin of activity in these systems, photon-based
control methods such as optical trapping and photon nudging 
are clearly applicable only if the motion of the particle can be actuated by means of photon induction. 
This actuation may occur via the thermophoretic effect, or by photophoretic or radiation-pressure forces \cite{ashkini}.
For instance, it has been demonstrated that metal-coated Janus particles show activity due to thermophoresis when
they are illuminated with  a strong laser light \cite{jia}.  Another example are Janus particles of the same type immersed in a binary solution with lower critical point.
This allows for local phase separation and, consequently, diffusiophoretic motion
with much lower laser intensities \cite{volpe2}. In the present work, we focus on those active particles which can be controlled by laser light. 

A simple model for the real particles of interest are so-called {\em active Brownian particles}. An active Brownian particle 
moves with constant velocity $v_0$ (motility), where the 
direction of this velocity changes in course of time due to 
 rotational noise. Most of the calculations in the present study are based on a fully two-dimensional version of an active Brownian particle 
 where both, the translations and the rotations, are restricted to say, the $x$-$y$ plane. The Langevin equations for this
 "disk-like" active particle are given by
\begin{eqnarray}
\dot{x}(t)&=&v_0 \cos\phi(t)-\nabla_x U(x,y) +\sqrt{2D_T} \xi_{T,x}(t) \nonumber \\
 \dot{y}(t)&=&v_0 \sin\phi(t)-\nabla_y U(x,y) +\sqrt{2D_T} \xi_{T,y}(t) \nonumber \\
\dot{\phi}(t)&=&\sqrt{2D_R} \xi_R(t),
\label{1}
\end{eqnarray}
where $x(t)$ and $y(t)$ are the components of the two-dimensional vector $\textbf{r}(t)=\left(x(t),y(t)\right)$, and $\phi(t)$ describes
the angle of the two-dimensional orientation (unit) vector $\hat{\bf e}(t)=\left(\cos\phi,\sin\phi\right)^T$ 
relative to the $x$-axis.
Further, $U(x,y)$ is the external potential, $D_T$ and $\xi_{T,x}(t)$,  $\xi_{T,y}(t)$ are the translational
diffusion constant and noise terms, respectively,  and $D_R$ and $\xi_R(t)$ relate to rotational diffusion. 
All noise terms here are independent, and each is considered to be Gaussian white noise with zero mean, i.e 
$\langle \xi(t) \rangle=0$, and  $\langle \xi(t_1)\xi(t_2) \rangle=\delta(t_1-t_2)$. 
In section~\ref{sec:spherical} we additionally consider a "spherical" active Brownian particles whose 
translational motion is still restricted to the $x$-$y$-plane (e.g., by some sort of confinement)
whereas the noisy rotational motion is three-dimensional.

An equivalent representation of the two-dimensional model in eqn.~(\ref{1}) is given by the Smoluchowski equation \cite{sevili}
\begin{equation}
\partial_t \psi= \lbrace\nabla\cdot ( \nabla U-v_0 \hat{\bf e}(t)+ D_T\nabla) + \partial^ 2_\phi \rbrace \psi,
 \label{fokki}
\end{equation}
where $\psi(\textbf{r},\phi,t)$ is the probability density function (PDF) of the particle position and the angle $\phi$.
Equation~(\ref{fokki}) is formally equivalent to the Smoluchowski equation for a passive Brownian particle 
with an additional force $v_0 \hat{\bf e}$. 

Calculating from eqn~(\ref{fokki}) the mean position of the active particle with the initial conditions $x(0)=0$, $y(0)=0$ and $\phi(0)=0$, one obtains 
\begin{eqnarray}
\langle x(t) \rangle &=& v_0 \tau_R [1-\exp(-\frac{t}{\tau_R})]\neq 0 \nonumber \\
\langle y (t)\rangle &=& 0
 \label{mean}
\end{eqnarray}
with $\tau_R=1/{D_R}$ being the relaxation time for rotational diffusion \cite{reva}. Equation~(\ref{mean}) 
indicates that the active Brownian particle performs a persistent motion in $x$-direction (due to the initial condition for $\phi$) before its direction is randomized. This unique 
effect is absent for passive particles.

 \section{Delayed feedback control }
In this section we propose a delayed feedback control strategy for steering an active particle
in the framework of two methods based on laser light.
\subsection{Optical trapping   }
\label{s2}
 The conventional optical trapping method without feedback control \cite{ashk} is based on a strong laser beam which acts like a  ''tweezer''. This laser tweezer 
restricts the random motion due thermal fluctuations by introducing 
 a confining potential. Apart from particle localization, the tweezer
provides the possibility to move the particle by moving the laser beam.
 However, tuning the laser intensity is not trivial: If the laser beam is not sufficiently strong to sharply localize the 
 particle, there is the possibility of losing the particle 
 while translating the laser beam position. This has to be balanced with the fact that a too strong intensity can damage the particle. While 
 these considerations apply already to passive particles, 
trapping of active particles can be even more involved since, upon switching on the laser beam, they can transform 
the received energy into directional motion.

However, when the laser beam is much larger than the particle size, the trapping
effect becomes significant only when then particle reaches the border of laser spot.  Within the spot, the motion of the particle 
consists of large free displacements due to the activity of the particles. Experimentally, it has been
demonstrated that such a set-up  could be provided by a defocused laser beam\cite{jiang}. The mean squared displacement (MSD)
of particles in a defocused laser beam has been experimentally  
shown to have a ballistic regime due to the directed motions of ABPs followed by a crossover to  normal diffusion at times longer
than the rotational relaxation time. Only at very long times, the MSD reaches 
a plateau due to the trapping effect in the laser beam. 

 In what follows we propose a delayed feedback control which could be coupled to the aforementioned optical trapping method in order 
 to localize and steer an active particle. 

 \subsubsection{Strategy of  control }
 In the following, we aim at developing a control mechanism by which one
 can guide the active particle from position $A$ to position $B$. 
 We assume both $A$ and $B$ to lie on the $x$-axis with $x_B>x_A$, specifically
 $x_B=x_A+L$.
 
We recall that the irradiation of a laser beam has two different impacts on the motion of a (photo-sensitive) active particle. 
On the one hand, it increases the mobility by creating a temperature gradient around the particle (self-thermophoresis) \cite{jiang}. 
On the other hand, it leads to a two-dimensional trapping of the particle  \cite{sato}. 
Here we aim at combining these two effects in order to
restrict the two-dimensional random motion of the particle with its three
degrees of freedom (i.e., $x$, $y$, $\phi$) to a quasi-one dimensional motion with a preferred direction. 
Our proposal for such a control process consists of  two steps: first,  restricting the motion in quasi-one dimension by an optical trap and second, breaking the symmetry of 
 motion by adapting the intensity.  
 
 In order to restrict the particle motion along one dimension, say $x$, one needs 
to enhance the trapping effect in $y$-axis. At the same time, the particle should be able to freely move in $x$-direction. 
This could be realized with a laser beam (with a waist being a few times the particle size to allow for limited free motion), whose center
moves with the particle, yet \textit{only} along the $x$-axis. 

Let us now construct the corresponding potential:
In general, a particle at position $\textbf{r}=(x,y)$  in an optical trap  located at $\textbf{r}_0=(x_0,y_0)$ experiences 
an approximately harmonic potential of the form 
\begin{equation}
U(x,y)=\frac{1}{2}\eta \left((x- x_0)^2+(y-y_0)^2\right),
\end{equation}
 where $\eta$, the spring constant or stiffness of the trap, depends on the laser intensity. 
 
 We recall that we want the particle to move from $\textbf{r}_A=(x_A,0)$ to $\textbf{r}_B=(x_B,0)$.
 We therefore set the $y$-component of the trap position $\textbf{r}_0$ to zero. Further we assume
 that there is a delay time $\delta$ between monitoring the position of particle at time $t$ and driving the laser along the $x$-axis.
 The "control" potential then has the form
\begin{equation}
 U_c(x,y,t)=\frac{1}{2}\eta \left((x(t)- {x}(t-\delta))^2+(y(t))^2\right).
\label{4}
\end{equation}
For small delay times (and thus, small differences $x(t)- {x}(t-\delta)$),
the trapping effect in $x$-direction is therefore much weaker than that in $y$-direction, and for $\delta\rightarrow 0$, the  particle feels
no trapping in $x$-direction at all. 

From eqn.~(\ref{4}), the control force acting on the particle follows as
\begin{eqnarray}
 \textbf{F}_c(t)&=& -\nabla U_c(x,y,t)\nonumber\\  
  &=&- \eta [\left(x(t)- {x}(t-\delta)\right)\hat{\bf i}+y(t)\hat{\bf j}],
  \label{amu}
\end{eqnarray}
where $\hat{\bf i}$ and $\hat{\bf j}$ are the unit vectors in $x$- and $y$-directions, respectively.   
The control expressed by eqn.~(\ref{amu}) prevents free motion of the particle in $y$-direction
and thus creates a "channel" along the 
$x$-direction.
After some time, the particle will indeed reach its destination ($B$) on the $x$-axis just by random motion. However, this purely diffusive mechanism can be improved.  

\begin{figure}
\scalebox{0.2}{\includegraphics{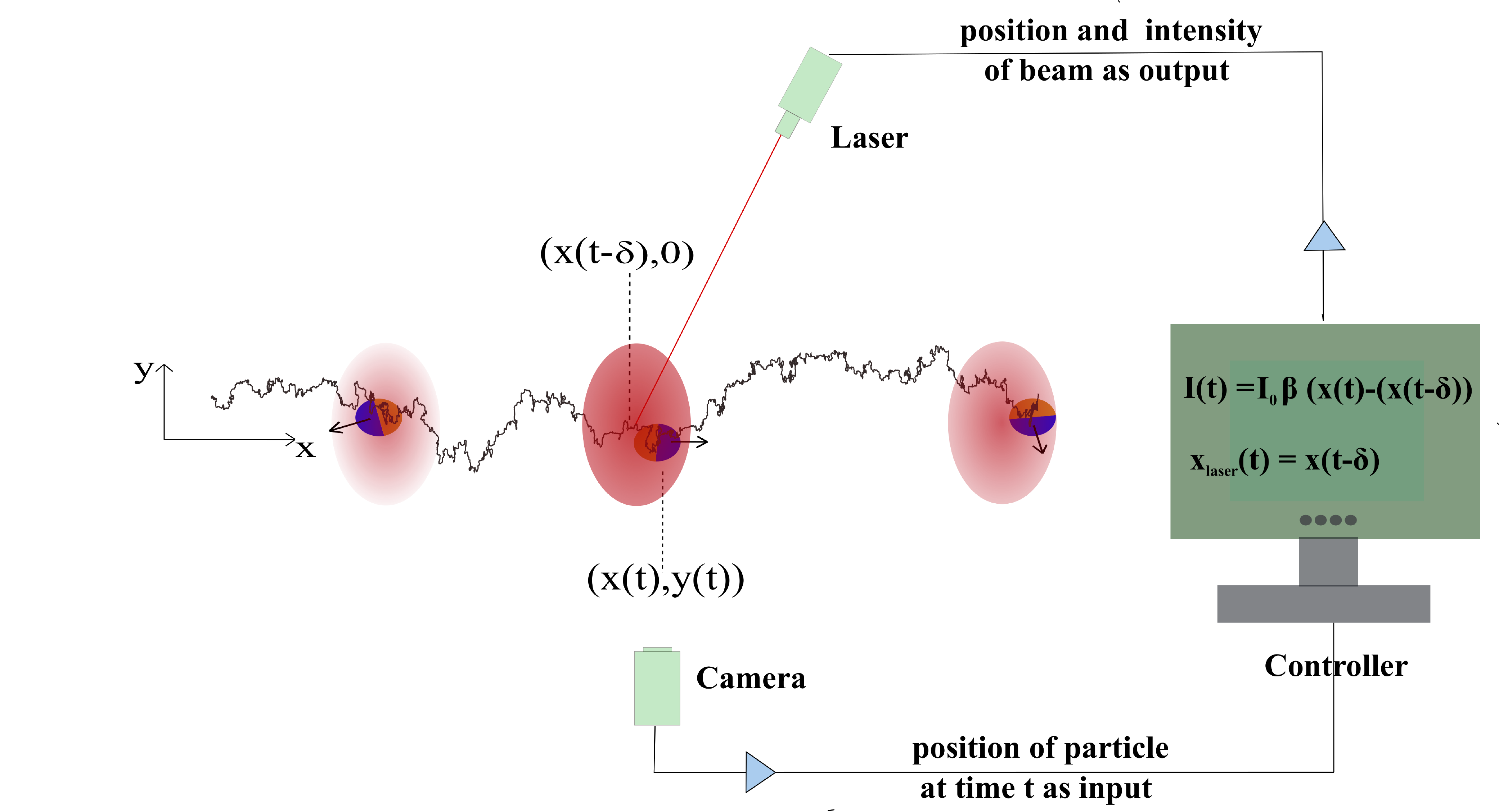}}
\caption{Schematic explanation of the control strategy involving optical trapping from a theoretical perspective. The 
intensity of the laser beam, which controls the motility of the particle, is a function of 
its displacement during the delay time (see eqn.~(\ref{mom})). If the particles is heading towards its destination, i.e., in positive
$x$-direction), the intensity is enhanced as indicated
by the deeper red color.}
\label{sketch}
\end{figure}

 To this end, the next step of our control strategy is to break the symmetry of the quasi-one dimensional motion along the $x$-axis in favor of our desired direction. 
 For this, we propose to
  modify the intensity of the  applied laser intensity depending on the previous position of the particle (see Fig.~\ref{sketch} for a sketch). This idea is based on the fact that the laser intensity determines both, 
  the motility ($v_0$) of the active particle and the stiffness ($\eta$) of the harmonic (optical) trap.
  
 Consider the difference between the particle's position in $x$-direction at time $t$, $x(t)$, and the corresponding position at the earlier time $t-\delta$.
 If the displacement $x(t)- x(t-\delta)$ is positive, 
 the particle
 is most likely heading in the desired direction towards its destination at point $B$ (with $x_B>x_A$).  Under this condition, 
 we increase the intensity in order to increase the motility. Likewise, for negative $x(t)- x(t-\delta)$ we decrease the intensity
  such that particle motion in the "wrong" direction is hindered. 
 Specifically, we assume a linear modification of the laser intensity described by
   \begin{equation}
I(t,\delta)= I(x(t)-x(t-\delta))= I_0\left(1+\beta \left(x(t)-x(t-\delta)\right)\right),
  \label{mom}
 \end{equation}
 where $\beta$ is a control parameter which determines the strength of symmetry breaking
 and has the dimension of an inverse length. Specifically, $\beta=0$ implies no symmetry-breaking whereas for $\beta>0$, displacements in the desired direction are supported
 by adapting the laser intensity. Clearly, the laser intensity should always remain positive. Therefore, $\beta$ has to be chosen such that the expression in the brackets in eqn.~(\ref{mom}) is always positive, that is, 
 $\beta (x(t)-x(t-\delta)  )  >  -1$.  The relation between  $\beta$ and   $\delta$ will become more clear  in the next section, where we apply a small-delay approximation

We now consider the resulting effect on the motility. 
 Experimental reports indicate  a linear relationship between motility and the laser intensity \cite{chicos2}, $v(t) \propto I(t)$, for moderate intensities. 
 Combining this with the above {\em ansatz} for the intensity, eqn.~(\ref{mom}), the
 motility at time $t$ for a given $\delta$ becomes
 \begin{equation} 
 v(t,\delta)=v_0 \left(1+\beta \left(x(t)-x(t-\delta\right)\right),
 \label{mm}
 \end{equation} 
where $v_0$ is the bare motility appearing in eqn.~(\ref{1}).
Finally, we take into account that the intensity of the laser beam changes also 
the stiffness of the laser trap, $\eta$. Assuming again a linear relationship \cite{stiff}, i.e. $\eta  \propto I$, we can write 
 \begin{equation} 
 \eta(t,\delta)=\eta_0 \left(1+\beta \left(x(t)-x(t-\delta)\right)\right).
 \label{ms}
 \end{equation} 

 One may note that the above considerations do not take into account a spatial dependency of the intensity and thus, the motility and stiffness, inside the trap.
We have indeed neglected such a dependency assuming 
that the laser beam is much bigger than the particle size ("defocused laser beam").  In this situation, the particle displacement during the delay time is so small 
that one may safely assume the intensity at any point in the optical trap to be equal. 
This assumption, however, is not crucial for our approach; one could easily include a spatial dependence as well.
 
We are now in the position to construct the equation of motion in presence of delayed feedback. To this end we start from the conventional
equations of motion for an active Brownian particle, eqn.~(\ref{1}). First,  
we replace the constant motility $v_0$ by  the 
time-dependent motility given in eqn.~(\ref{mm}). Second, we replace the derivative of the potential $U(x,y)$
by the control force given in eqn.~(\ref{amu}), where the spring constant 
is now given by eqn.~(\ref{ms}). 
 With these steps we arrive at
\begin{eqnarray}
\dot{x}(t)&=& v_0\left(1+\beta \left(x(t)-x(t-\delta)\right)\right) \cos\phi(t) \nonumber\\
          & &- \eta_0 \left(1+\beta \left(x(t)-x(t-\delta)\right)\right)\times\left(x(t)- x(t-\delta)\right)\nonumber \\ 
          & &  + \sqrt{2D_T} \xi_{T,x}(t), \nonumber \\
\dot{y}(t)&=&v_0\left(1+\beta \left(x(t)-x(t-\delta)\right)\right) \sin\phi(t)  \nonumber \\ 
          & &-\eta_0  \left(1+\beta \left(x(t)-x(t-\delta)\right)\right)\times  y(t)  \nonumber \\
          & &  +\sqrt{2D_T} \xi_{T,y}(t), \nonumber \\ 
\dot{\phi}(t)&=&\sqrt{2D_R} \xi_R(t).
\label{bl}
\end{eqnarray}

From a mathematical perspective, eqns.~(\ref{bl}) represent a set of coupled stochastic delay-differential equations. 
Treating such systems is generally a challenging task: For example, the delay can induce new dynamical features
such as spontaneous oscillations \cite{tsimri,hongi} not seen in the Markovian case. Moreover, the transition
towards a Fokker-Planck description is significantly more involved (see, e.g., \cite{loosi}).

In the present study we proceed with the theoretical description by assuming
that the delay time, $\delta$, is much shorter than the rotational relaxation time, $\tau_R$. In other words,
there is only a very small displacement of the particle during the delay time. This assumption justifies 
a Taylor expansion of $x(t-\delta)$ around $\delta=0$, i.e., $x(t-\delta)\simeq x(t)-\delta \dot{x}(t)+{\cal O}(\delta^2)$.
Substituting the Taylor expansion into eqns.~(\ref{bl}), neglecting all terms in $\delta$ beyond the linear one and solving the resulting equations
with respect to $\dot x$ and $\dot y$, we obtain the non-delayed (and thus, Markovian) equations
\begin{eqnarray}
\dot{x}(t)&=&   v_0 \cos\phi+\beta \delta  v_0^2 \cos^2\phi- \eta_0 v_0 \delta  \cos\phi\nonumber\\
          & &+\left(1+\beta \delta  v_0 \cos\phi- \eta_0 \delta\right)\sqrt{2D_T} \xi_{T,x}, \nonumber \\ 
\dot{y}(t)&=&  v_0 \sin\phi+ \beta \delta v^2_0 \sin\phi \cos\phi \nonumber \\
          & & -\eta_0 y \left(1+\beta \delta v_0 \cos\phi+\beta \delta \sqrt{2 D_T} \xi_{T,y} \right) \nonumber \\
          & & + \beta \delta v_0 \sin\phi \sqrt{2 D_T} \xi_{T,x}+\sqrt{2D_T} \xi_{T,y}, \nonumber \\
\dot{\phi}(t)&=&\sqrt{2D_R} \xi_R(t).
\label{fl}
\end{eqnarray}

We recall our argumentation in the previous section (after eqn.~(\ref{mom})) regarding the limitation in choosing the  value of $\beta$. This restriction can now be formulated more precisely. Applying the small delay 
approximation to the right side of eqn.~(\ref{mom}), one obtains the condition $\beta \delta  \dot{x}(t)  >  -1$, which relates the $\beta$ to the delay time $\delta$. This implies that to keep the intensity positive, and noting
that the sign of  $\dot{x}(t)$ can become negative, large values of $\beta$ require small values of $\delta$. 
\subsubsection{Coarse-grained equations of motion}
\label{s3}
When steering a particle from position $A$ to $B$, we are mainly interested in two quantities: first, the average time $\tau_{arr}$ which the particle needs to 
arrive at the target position,
and second, a measure for the deviation between the (fictitious) straight motion towards the target and the actual path. 
Given that both  $A$ to $B$ are located on the $x$-axis, the deviation can be quantified through the root mean squared displacement in $y$-direction, $\langle y^2 \rangle$. 
In principle, one may calculate $\tau_{arr}$ and $\langle y^2 \rangle$
by direct integration of eqns.~(\ref{fl}). 
%
%
In this way, however, one takes into account the full trajectory which includes times smaller than the rotational relaxation time, $\tau_R$.
One would expect these small times to be essentially irrelevant for the long-time behavior which determines the quantities of interest.
In what follows, we therefore derive coarse-grained equations of motion focusing on the translational dynamics alone. That is, we aim at integrating out the rotational variable.
This allows us to obtain the desired quantities at longer time scales beyond the rotational relaxation time.

Our coarse-graining procedure closely follows that proposed by Mijialkov \textit{et al.} \cite{Mijalkov} and Leyman {\em et al.}  \cite{leyman2018tuning}
(see the Appendices of these references for details). 
We start by considering the limiting case of
eqns.~(\ref{fl}), where the rotational relaxation time $\tau_R$ and the delay time $\delta$ are both very small, but their ratio is finite and tends to zero, i.e., $\delta/\tau_R\rightarrow 0$.
Formally, this is done by introducing a small parameter $\epsilon$ such that 
$\delta= c \epsilon$ and $\tau_R =k\epsilon$, with $c$ and $k$ being constants and $\epsilon\rightarrow 0 $. 
In order to keep the particle displacement significant for small $\tau_R$, we define $u= \sqrt{\tau_R} v_0$. 
Inserting these definitions into eqns.~(\ref{fl}) and keeping only the leading forms i.e., those of order $\epsilon^{-\frac{1}{2}}$ and unity, we obtain 
\begin{eqnarray}
\dot{x}(t)&=&  \frac{u}{\sqrt{k\epsilon}} \cos\phi+\beta \frac{c}{k} u^2 \cos^2\phi + \sqrt{2D_T} \xi_{T,x}, \nonumber \\ 
\dot{y}(t)&=&  \frac{u}{\sqrt{k\epsilon}} \sin\phi+ \beta \frac{c}{k} u^2 \cos\phi \sin\phi  -\eta_0 y+\sqrt{2D_T} \xi_{T,y}, \nonumber \\
\dot{\phi}(t)&=&\sqrt{\frac{2}{k \epsilon}} \xi_R(t).
\label{fl1}
\end{eqnarray}
The PDF $\tilde\psi(\textbf{r}, \phi,t)$ corresponding to eqns.~(\ref{fl1}) (i.e., the rescaled Langevin equations in small-delay approximation) obeys the backward Kolmogorov equation
\begin{equation}
\partial_t \tilde\psi= \left \{ \frac{1}{k \epsilon}  \partial^ 2_\phi+ (\frac{u}{\sqrt{k\epsilon}}+\beta\frac{c}{k} u^2 \cos\phi)  
 \hat{\bf e}\cdot\nabla-\eta_0 y \partial_y + 
D_T \Delta \right \} \tilde\psi,
 \label{backi}
\end{equation}
where $\Delta$ and $\nabla$ refer (only) to spatial derivatives,
i.e., $\Delta = \partial^2/\partial x^2+\partial^2/\partial y^2$ and  $\nabla= (\partial/\partial x) 
\hat{\bf i}+(\partial/\partial y) \hat{\bf j}$.
  Here we are interested in a {\em reduced} backward Kolmogorov equation which describes the coarse-grained probability distribution function, $\psi_0(x,y,t)$, 
which is independent of the rotational degree of freedom. To this end we employ the multiscale
expansion method \cite{multiscale}. Specifically, we expand $\tilde\psi$ in powers of $\sqrt \epsilon$ as 
\begin{equation}
\tilde\psi= \psi_0(\textbf{r},t)+ \sqrt \epsilon \psi_1(\textbf{r},\phi,t)+\epsilon \psi_2(\textbf{r},\phi,t)+\ldots .
\label{psiexp}
\end{equation}
We further note that eqns.~(\ref{backi}) may be written as 
\begin{equation}
\partial_t \tilde\psi= \left \{ \frac{1}{ \epsilon}  L_0+\frac{1}{\sqrt{\epsilon}} L_1+ L_2 \right \} \tilde\psi
 \label{backi1}
\end{equation}
where the operators $ L_0$, $L_1$ and $L_2$ are defined as
\begin{eqnarray}
L_0&=& \frac{1}{k}  \partial^ 2_\phi , \nonumber\\
L_1&=& \frac{u}{\sqrt{k}}  \hat{\bf e}\cdot\nabla, \nonumber\\
L_2&=& \beta\frac{c}{k} u^2 \cos\phi   \hat{\bf e}\cdot\nabla
-\eta_0 y \partial_y+ D_T \Delta. 
\label{L_def}
\end{eqnarray}
Inserting the {\em ansatz}~(\ref{psiexp}) into eqn.~(\ref{backi1}) and sorting the terms according to their order in $\sqrt\epsilon$, we obtain 
in order $(\sqrt\epsilon)^{-2}=\epsilon^{-1}$
\begin{equation}
L_0 \psi_0= \frac{1}{k}  \partial^ 2_\phi \psi_0= 0,
 \label{backi11}
\end{equation}
where the first part of the equation results from using the first member of eqn.~(\ref{L_def}). Similarly, 
in order $\sqrt\epsilon$ we find
\begin{equation}
L_1 \psi_0+ L_0 \psi_1= \frac{u}{\sqrt{k}} \hat{\bf e}\cdot\nabla\psi_0+ \frac{1}{k }  \partial^ 2_\phi \psi_1=0,
 \label{backi12}
\end{equation}
and finally, in order $(\sqrt\epsilon)^0=1$,
\begin{eqnarray}
L_1 \psi_1+ L_2 \psi_0+ L_0 \psi_2= \frac{u}{\sqrt{k}} \hat{\bf e}\cdot\nabla  \psi_1+ \nonumber\\
\left(\beta\frac{c}{k} u^2 \cos\phi  \hat{\bf e}\cdot\nabla-\eta_0 y \partial_y + D_T \Delta \right) \psi_0 \nonumber\\
+\frac{1}{k}  \partial^ 2_\phi \psi_2= \partial_t \psi_0.
 \label{backi13}
\end{eqnarray}
Equation~(\ref{backi11}) allows for a solution $\psi_0$ which contains a  linear term in $\phi$ and a constant. Here we select the constant solution
since the coarse-grained PDF should not depend on $\phi$, i.e.,  $\psi_0=\psi_0(x,y,t)$. 
Further, eqn.~(\ref{backi12}) implies that $\psi_1=\sqrt{k} u \hat{\bf e}\cdot{\nabla}\psi_0$ where we 
have used the fact that $\hat{\bf e}=(\cos\phi,\sin\phi)$. 
Finally, eqn.~(\ref{backi13}) can be rewritten as $L_0 \psi_2 = \partial_t\psi_0-L_1\psi_1-L_2\psi_0$. Formally
(see Ref.~\cite{multiscale} for a more mathematical discussion)
this implies that the function $u({\bf r},\phi,t)=\partial_t\psi_0-L_1\psi_1-L_2\psi_0$ is in the range of 
the operator $L_0$,  where  $L_0$ is an operator in $\phi$ and $u$ is considered as  a function slowly varying in $\phi$.   As a consequence \cite{multiscale}, $u$ has to be {\em orthogonal} to the null space (kernel) of the adjoint operator $L_0^{*}$ (where the null space is the set of functions $\rho$ for which $L_0^{*}(\rho)=0$). Here, $L_0$ is a self-adjoint operator, i.e.,
$L_0^{*}=L_0$, and the null space corresponds to a constant function in $\phi$, as argued already at the beginning of this paragraph. This yields the orthogonality relation 
\begin{eqnarray}
\int_0^{2\pi} d\phi \,u({\bf r},\phi,t)\times \rho &=&0,
\end{eqnarray}
or, equivalently, 
\begin{equation}
\partial_t\psi_0(x,y,t)= \frac{1}{2\pi} \int_0^{2\pi} d\phi \left( L_1\psi_1+L_2\psi_0\right)
\end{equation}
Substituting $\psi_1=\sqrt{k} u \hat{\bf e}\cdot{\nabla}\psi_0$, as obtained above,  in the integral and performing the integration, one reaches to the desired backwarded Kolmogorov equation for the
coarse-grained PDF $\psi_0$, that is,
\begin{equation}
\partial_t \psi_0(x,y,t)= \lbrace \beta \delta \frac{v_0^2}{2} \partial_x- \eta_0 y \partial_y + 
(D_T+\tau_R \frac{v_0^2}{2}) \Delta \rbrace \psi_0(x,y,t).
 \label{backf}
\end{equation}
As a last step of our coarse-graining strategy, we note that eqn.~(\ref{backf}) corresponds to the following set of (Markovian) Langevin equations
for the variables $x$ and $y$,
\begin{eqnarray}
\dot{x}(t)&=&  \beta \delta \frac{v_0^2}{2} + \sqrt{2D_T+ \tau_R v_0^2} \xi_{T}^{(1)}, \nonumber \\ 
\dot{y}(t)&=&  -\eta_0 y+ \sqrt{2D_T+ \tau_R v_0^2} \xi_{T}^{(2)},
\label{flf}
\end{eqnarray}
where $\xi_{T}^{(1)}$, $\xi_{T}^{(2)}$ are again Gaussian white noises.

 A "pedestrian" proof of the drift term in the equation for $\dot{x}$ and the friction term in the equation for  $\dot{y}$ in equations above  can be  done by looking at the long time asymptotic behavior of each term in eqns.~(\ref{fl}) or eqns.~(\ref{fl1}).
 At very long times, the particle visits all the orientations with the same probability, that is the angular probability density  is a constant $P(\phi,t)=\frac{1}{2\pi}$. This corresponds to a unweighted average over $\phi$.
 Thus, terms linear in $\sin \phi$ and $\cos \phi $ will not effectively drive the particle. The
 term $\cos^2 \phi$
 in the  equation for $\dot{x}$ will, however, remain positive and its average of $1/2$  leads  to a drift term as $\beta \delta v_0^2/2$.  In the equation for the $y$ component, the $-\eta_0 y$ term is 
 independent of the particle orientation. This leads to the friction term.

 In conclusion, inspecting the first member of the coarse-grained Langevin eqns.~(\ref{flf}), we find that the active particle effectively feels a constant driving force of magnitude  $\beta \delta v_0^2/2$
in positive $x$-direction, i.e., towards its destination.
As a consequence, the average position in $x$-direction at time $t$ is given by
$\langle x(t) \rangle =
\beta \delta v_0^2/2 t$. We recall that the separation between the target position $x_B$ and
the initial position $x_A$ is given by $L$. From this, we obtain 
the mean arrival time
\begin{eqnarray}
{\tau}_{arr}= \frac{2 L}{v_0^2 \beta \delta} .
\label{arrive}
\end{eqnarray}

The $y$-component of the particle position (see the second member of eqns.~(\ref{flf})) is, however, described by an Ornstein-Uhlenbeck process.
From this one can calculate the mean $y$-position,
\begin{equation}
\langle y(t) \rangle =0
\end{equation}
and the mean squared displacement
in $y$-direction,
\begin{equation}
\langle y^2 (t) \rangle = \frac{D'}{\eta_0}\left(1-\exp(-2\eta_0 t)\right)
\label{y2}
\end{equation}
with the renormalized diffusion constant $D'=2 D_T + \tau v_0^2$. From eqn.~(\ref{y2}) we finally obtain the long-time limit
$\langle y^2 \rangle _{t\rightarrow  \infty}  =D'/\eta_0$.
 \subsubsection{Simulation}
\label{s4}
To check the predictions of our coarse-grained analytical theory, particularly the result for the mean arrival time (see eqn.~(\ref{arrive})), we have performed numerical simulations
of the full, delayed stochastic equations of the motion given in eqns.~(\ref{bl}). 
In these simulations, the units of time and length were set to 
the delay time, $\delta$, and the size of the particle, $\sigma$, respectively. 
To comply with a realistic experimental situation \cite{hauf}, the rotational relaxation time was chosen to be $65$ times longer than the delay time, and
the translational noise was neglected, i.e., $D_T=0$. The stiffness of the trap was set to $\eta_0=0.1 \sigma \delta^{-2}$.
Finally, the motility $v_0$ was set to  $1 \sigma/\delta$. This is sufficiently small such that in one unit of time, the particle stays in the spot created by the laser. 

\begin{figure}
\scalebox{0.33}{\includegraphics{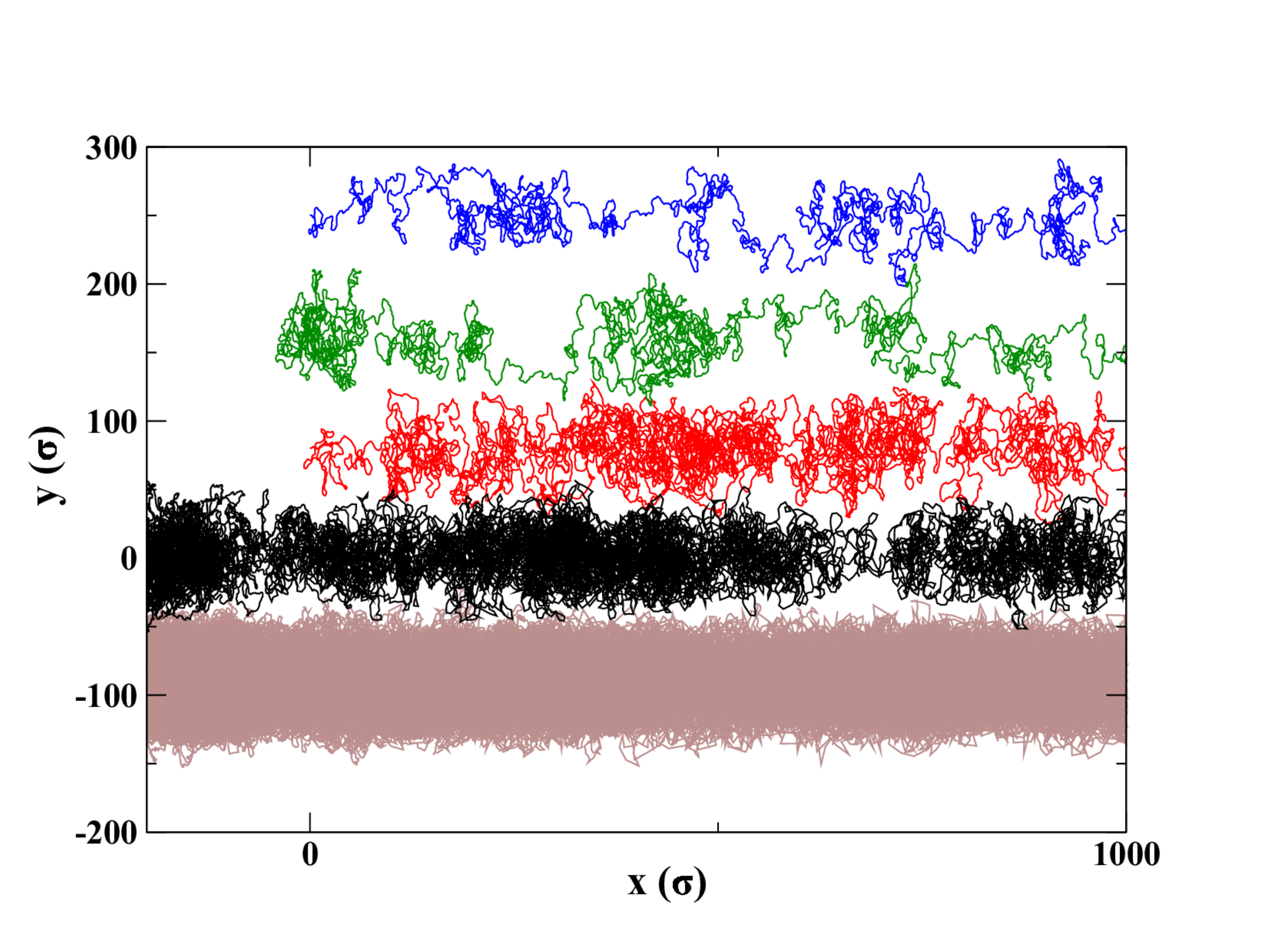}}
\caption{Exemplary particle trajectories in the $x$-$y$ plane according to eqns.~(\ref{bl})
for $\beta\sigma=0$ (brown), $5$ (black), $10$ (red), $15$ (green) and
$20$ (blue). For better visibility, the data corresponding to different $\beta\sigma$ are shifted by $\Delta y_A/\sigma=80$.
The other parameters are set to $\tau_R=65\delta$, $v_0=1\sigma/\delta$, and $\eta_0=0.1 \sigma \delta^{-2}$.}
\label{et}
\end{figure}

\begin{figure}
\scalebox{0.33}{\includegraphics{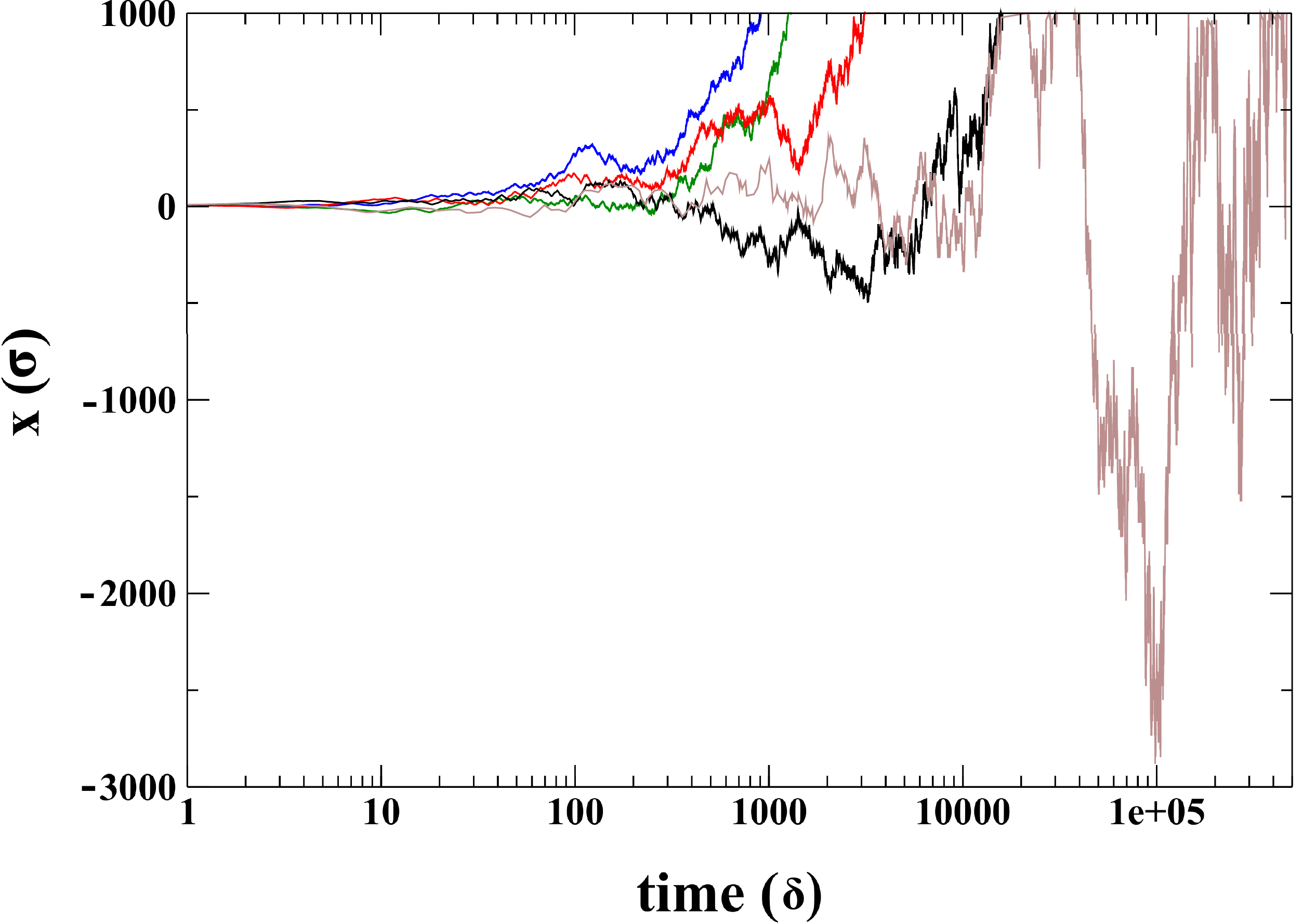}}
\caption{The $x$-component of the particle position as function of time according to eqns.~(\ref{bl}) for $\beta\sigma=0$ (brown), $5$ (black), $10$ (red), $15$ (green) and
$20$ (blue). The other parameter are chosen as in Fig.~\ref{et}.}
\label{2et}
\end{figure}

In figures~\ref{et} and \ref{2et} we present exemplary particles trajectories, first, in the $x$-$y$ plane (Fig.~\ref{et}) and second, in $x$-direction as function of time (Fig.~\ref{2et}).
The particle moves from the starting point at $\textbf{r}_A=(0,y_0)$ to the target position 
$\textbf{r}_B=(1000 \sigma, y_0$), where the different values of $y_0$ are solely used to separate different trajectories. 
The shown trajectories differ by the parameter $\beta\sigma$, where we recall that $\beta$ (which has the dimension of an inverse length) controls the strength of symmetry breaking in $x$-direction.
For better visibility, we focus on the range $y>-200\sigma$.
%
%

We start by considering the case $\beta\sigma=0$.
In this case, the laser intensity and thus, the motility and stiffness are always constant 
(see eqns.~(\ref{mom}), (\ref{mm}), and (\ref{ms}))
and there is no symmetry breaking in $x$-direction. The impact of the control then
reduces to the trapping in $y$-direction. The latter effect is clearly visible from Fig.~\ref{et}, where
the trajectory for $\beta\sigma=0$ appears like a densely filled "stripe". 
However, this shape of the trajectory also implies that the particle moves randomly to the right and left, that is,
there is no bias.
The latter point is even better seen in Fig.~\ref{2et}, which shows the $x$-position as function of time. Clearly, for $\beta\sigma=0$ there
is no preference for negative or positive values of $x$. 

Increasing the parameter $\beta\sigma$ from zero, the trapping in $y$-direction continues to be effective. This is seen from
Fig.~\ref{et}, illustrating that the area explored by the particle in $y$-direction stays essentially constant compared to $\beta\sigma=0$.
Importantly, however, the motion in $x$-direction becomes more and more directed towards positive values.
In more detail, at the lowest nonzero value considered ($\beta\sigma=5$), particle motion in negative direction is still significant. Closer inspection
reveals that the trajectory (in $x$-$y$-plane) here consists of large loops which slowly move towards positive $x$-values. 
For larger values of $\beta\sigma$ the symmetry breaking is more significant and  displacements in negative $x$ direction become progressively shorter (see, e.g., the case
$\beta\sigma=20$). 

These effects are even better visible in Fig.~\ref{2et}, showing clearly the importance of the symmetry breaking in $x$-direction to push the particle into the right direction.
From a mathematical point of view, this becomes understandable when we take a look at the first member of eqns.~(\ref{flf}). For small values of $\beta$, the effective noise described by the last term
competes with the drift term, the latter being proportional to $\beta$. Consequently, the particle experiences significant fluctuations im $x$-direction. These fluctuations
become more and more restricted when the drift term is enhanced by increasing $\beta$. 
  
\begin{figure}
\scalebox{0.33}{\includegraphics{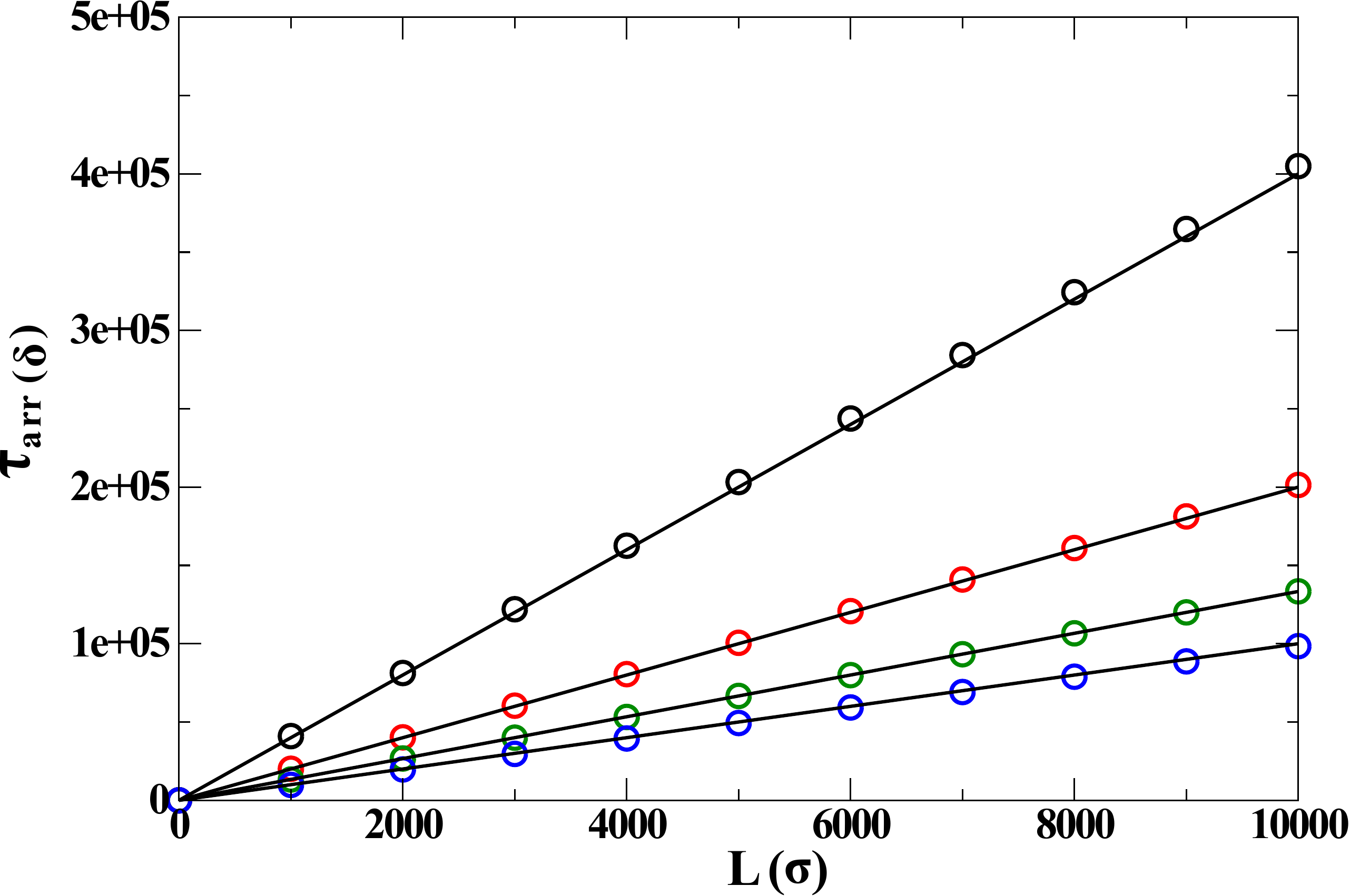}}
\caption{Mean arrival time of the particle as a function of the distance 
between starting point and target for $\beta\sigma=5$ (black), $10$ (red), $15$ (green) and $20$ (blue).
Symbols correspond to averaged numerical results from eqns.~(\ref{bl}), whereas
black lines represent plots of eqn.~(\ref{arrive}). }
\label{3et}
\end{figure}

We now turn to the mean arrival time, which the particle requires to reach its target. To obtain numerical results, we have performed $10^3$ simulation runs for each value of $\beta\sigma$.
The averaged numerical data are shown in Fig.~\ref{3et}, which also includes the analytical prediction from the coarse-grained theory (see eqn.~(\ref{arrive})).  

At the smallest value 
considered, $\beta\sigma=5$, we observe small deviations between theory and simulation data. The reason is that in this weakly controlled 
situation, fluctuations in $x$-direction are non-negligible, as already explained above.
In contrast, we observe excellent agreement between theory and simulations at larger values of $\beta\sigma$. This agreement is due to the fact
that, within our analytical theory, larger values of $\beta$ correspond to a larger drift term
in the (coarse-grained) equation for the $x$-component, see eqn.~\ref{flf}. The impact of the noise term (second term) then becomes negligible.
One should note, however, that even in the numerical simulations the control parameter $\beta$ cannot be chosen arbitrarily large: the reason
is that the intensity, which depends on $\beta$ and on the displacement (see eqn.~(\ref{mom})), has always to remain positive, yielding the condition
 $1+\beta \left(x(t)-x(t-\delta)  \right )  \geq 0$. For negative displacements, this imposes an upper limit for $\beta$.
 
  Finally, it is worth to note that essentially the same efficiency of our control strategy (quantified through the mean arrival time)
could be achieved if we kept the stiffness of the trap, $\eta$, constant. 
Indeed the main effect of a constant stiffness is that the width of the channel in 
$y-$direction becomes constant. However, test calculations showed that this is essentially irrelevant for the resulting mean arrival time. 
The underlying reason can be seen from the original (delayed) equations of motion, eqn.~(\ref{bl}). These equations reflect that the stiffness
enters the dynamics of the $x$-coordinate only to second order in the displacement (in contrast, the dynamics of the $y$-coordinate is affected already in linear
order). Therefore, the mean arrival time for steered motion in $x$-direction (for which our proposal has been formulated),
is only weakly affected by the time dependence of the stiffness. If we considered more refined measures of efficiency,
such as the degree of fluctuations in perpendicular ($y$-) direction, these would certainly be more sensitive.

\subsubsection{Role of rotational noise}
\label{sec:spherical}
So far we have applied our control mechanism to a "disk-like" active particle where not only the translational motion,
but also the possible orientations (and thus, directions of self-propelled motion) are confined to a plane.
Given this restriction it is an interesting question whether the control would work as well for the somewhat more realistic model of a 
"spherical" active particle, whose motion is still two-dimensional, but whose orientation can explore the full three-dimensional space. 
That is, the orientation vector $\hat{\bf{e}}$ can point along any direction on the (unit) sphere, $\hat{\bf{e}}=\left(\sin\theta\cos\phi,\sin\theta\sin\phi,\cos\theta\right)$,
with $\phi$ and $\theta$ being the polar and azimuthal angle, respectively. We refer to this model henceforth as "3$d$", where $d$ is the dimension of rotational motion.
From a physical point of view, this 3$d$ model describes, e.g., an active colloid which is resting on the bottom
of a container, or a confined particle which motion in $z$-direction is restricted by walls \cite{selm,hydrodynamic,zottl2016emergent}. 
Assuming again a two-dimensional self-propulsion of the form $v_0(e_x,e_y)$, the corresponding Langevin equations
read
\begin{eqnarray}
\dot{x}(t)&=&v_0 \sin\theta(t)\cos\phi(t)-\nabla_x U(x,y) +\sqrt{2D_T} \xi_{T,x}(t) \nonumber \\
 \dot{y}(t)&=&v_0 \sin\theta(t)\sin\phi(t)-\nabla_y U(x,y) +\sqrt{2D_T} \xi_{T,y}(t) \nonumber \\
\dot{\bm{e}}(t)&=&\sqrt{2D_R}\bm{e}(t)\times \bm{\xi_r}(t),
\label{eq:3D}
\end{eqnarray}
where $\bm{\xi_r}$ is a stochastic torque modeled by Gaussian white noise with zero mean and delta-like
correlation in time. The correlation function of the orientation vector is then given by \cite{dhont1996introduction} $\langle \bm{e(t)}-\bm{e(0)}\rangle= e^{-t/\tau_R} $ with $\tau_R$ being 
 the rotational relaxation time. We note that for rotational noise of dimension $d$, the relation between rotational relaxation time and rotational diffusion constant is given by
$\tau_R= 1/\left((d-1)D_R\right)$. For $d=3$ this yields $\tau_R= 1/(2D_R)$ (contrary to the $2d$, disk-like, case studied before, see eqn.~(\ref{mean}) below).
It is worth mentioning that the rotational noise in eq.~(\ref{eq:3D}) is of {\em multiplicative} character (with possible 
implications discussed , e.g., in Refs. [\cite{selm,hydrodynamic,zottl2016emergent}]).

In analogy to our procedure for the 2$d$ model (see section~\ref{s2}), we now
replace the constant motility $v_0$ by  the 
time-dependent motility given in eqn.~(\ref{mm}), and the derivative of the potential $U(x,y)$
eqn.~(\ref{amu}), utilizing eqn.~(\ref{ms}) for the spring constant. This yields the delayed Langevin equations
\begin{eqnarray}
\dot{x}(t)&=& v_0\left(1+\beta \left(x(t)-x(t-\delta)\right)\right) e_x(t) \nonumber\\
          & &- \eta_0 \left(1+\beta \left(x(t)-x(t-\delta)\right)\right)\times\left(x(t)- x(t-\delta)\right)\nonumber \\ 
          & &  + \sqrt{2D_T} \xi_{T,x}(t), \nonumber \\
\dot{y}(t)&=&v_0\left(1+\beta \left(x(t)-x(t-\delta)\right)\right) e_y(t) \nonumber \\ 
          & &-\eta_0  \left(1+\beta \left(x(t)-x(t-\delta)\right)\right)\times  y(t)  \nonumber \\
          & &  +\sqrt{2D_T} \xi_{T,y}(t), \nonumber \\ 
\dot{\bm{e}}(t)&=&\sqrt{2D_R}\bm{e}\times \bm{\xi_r}(t)
\label{bio1}
\end{eqnarray}

\begin{figure}
\scalebox{0.33}{\includegraphics{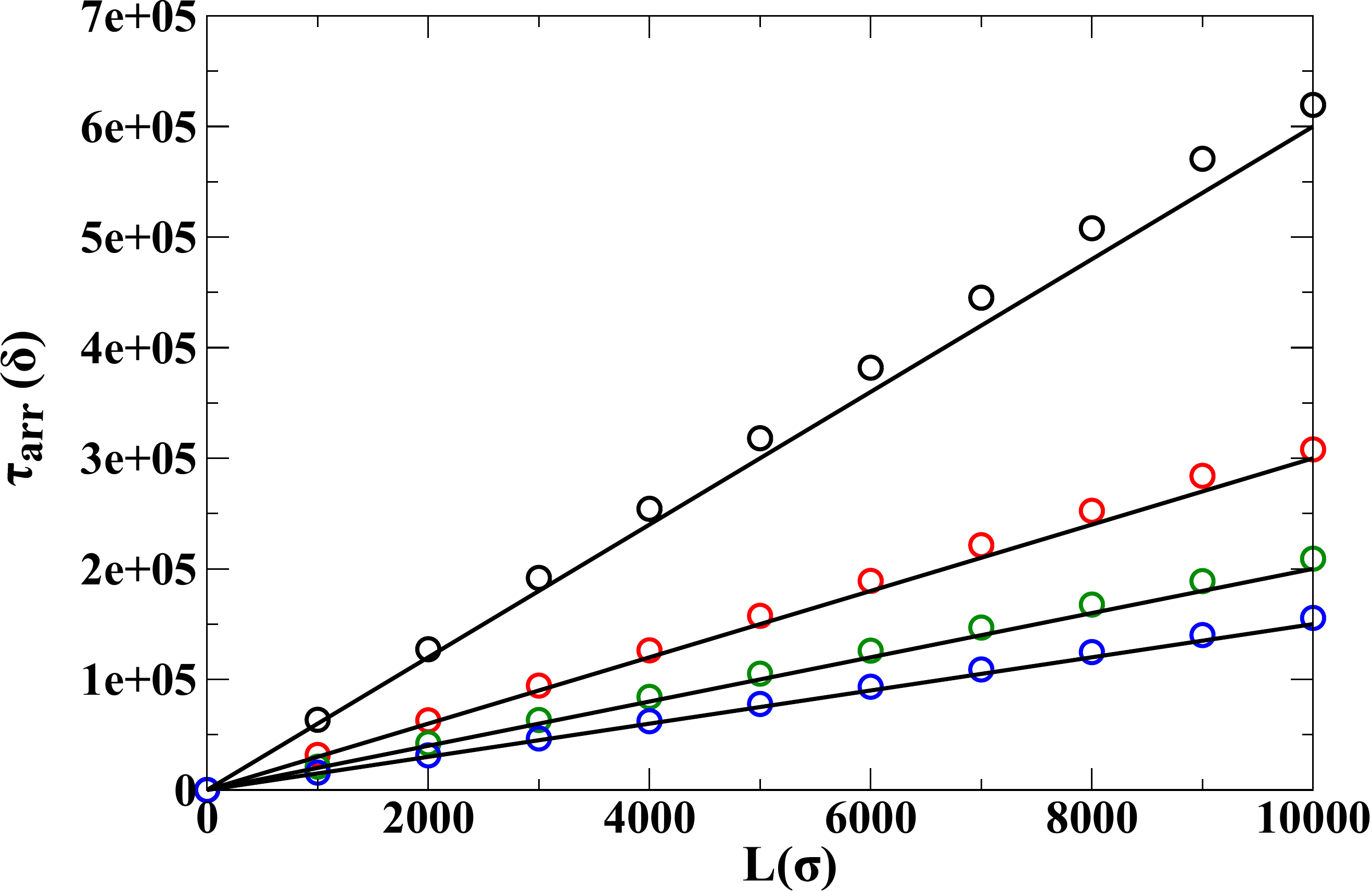}}
\caption{Mean arrival time of a confined active particle with three-dimensional rotational noise as function of the distance 
between starting point and target for $\beta\sigma=5$ (black), $10$ (red), $15$ (green) and $20$ (blue).
Symbols correspond to averaged numerical results from eqns.~(\ref{bio1}) [CHECK!!], whereas
black lines represent plots of the function ${\tau}_{arr}^{3d}$ defined in eq.~(\ref{arrive_3d}). }
\label{3d}
\end{figure}

To study the impact of the different character of rotational noise (as compared to the 2$d$ model considered before), we have performed
a set of numerical simulations similar to those described in the previous section, for control parameters $\beta\sigma=5$, $10$, $15$ and $20$.
Results for the mean time which a particles needs to move over a distance $L$ on the $x$-axis 
are plotted in Fig.~\ref{3d}. The data indicate again a {\it linear} dependence of the arrival time of the distance, consistent with
what we have seen in the case of two-dimensional rotational motion, see Fig.~\ref{3et}. However, closer inspection shows that
(within the errors arising from the noise terms in the equations of motion) the mean arrival times in the case of 3$d$ rotational noise
 are larger by almost fifty percent. This implies, in particular, that the prediction for $\tau_{arr}$ of 
our coarse-grained model, eqn.~\ref{arrive}, which gave a very good estimate for the 2$d$ situation (see Fig.~\ref{3et}),
does not properly describe the 3$d$ case.

To this end, it is helpful to have a closer look on the implications of the definition of the orientation vector. In the 3$d$ case,
the $x$ and $y$ components are given as $e_x=\sin\theta\cos\phi$ and $e_y=\sin \theta \sin\phi$, where the factor $\sin\theta$ makes
the difference to the 2$d$ case. From eqn.~(\ref{eq:3D}) or (\ref{bio1}) it can be seen that
this factor may also be regarded as a {\em prefactor} of the motility, $v_0$, suggesting the definition of a modified motility $v'_0=v_0 \sin \theta$. 
With this modified motility, equations~(\ref{bio1}) for $\dot{x}(t)$ and $\dot{y}(t)$ become identical to those in the $2d$ case, see eqn.~(\ref{bl}).
We now turn back to our earlier prediction of the mean arrival time, eqn.~\ref{arrive}, which involves an inverse quadratic dependence on the motility.
To apply this to the 3$d$ case, we suggest to replace $v_0^2$ by $\langle v'_0)^2\rangle=v_0^2\langle \sin^2\theta\rangle$. To estimate the average
we use a similar argument as we did within our "pedestrian proof" below eqn.~(\ref{flf}):
 At long times (i.e., large distances to the starting point), the angle $\theta$ explores all values in the interval $[0,\pi]$, with a weight being given by the (normalized) distribution
 is $P(\theta)= 2^{-1}\sin\theta$ (corresponding to a uniform sampling of the unit sphere).
 The average is therefore given as 
 $\langle (v'_0)^2\rangle=v_0^2\langle \sin^2\theta\rangle=v_0^2 \int_0^\pi d\theta P(\theta) \sin^2\theta=2 v_0^2/3$.
 Inserting this expression into eqn.~\ref{arrive}, we obtain
 \begin{eqnarray}
{\tau}_{arr}^{3d}= \frac{3 L}{v_0^2 \beta \delta}.
\label{arrive_3d}
\end{eqnarray}

Comparing the resulting data for the mean arrival times with the numerical ones, see Fig.~\ref{3d}, we find that the agreement is again remarkably good, similar as it was
in the 2$d$ case. Of course, one reason for the simple relation between the analytical results for the mean arrival time in the 2$d$ and 3$d$ case is that,
even for the 3$d$ situation, we still consider a motion along the $x$-axis. An analytical treatment for a path in two or three spatial dimensions would be much more involved.
Also, our treatment of the 3$d$ case neglects any frictional forces, which might be present for a real spherical particle confined by a wall in $z$-direction.

\subsection{Photon Nudging: Control Strategy }
\label{nudging}
While optical trapping has, generally, a wide range of applications, 
it also has drawbacks (e.g., destruction) and limitations in complex systems such as biological environment \cite{neumane}, even at low laser intensity.
This motivated us to examine the proposed idea to approximate the particle orientation in the context of the photon nudging
  method, where the laser intensity is typically even smaller and the laser is not continuously active.

More specifically, within the photon nudging method \cite{chicos2},  a focused laser beam of moderate intensity pushes the active particle along its 
heading direction. Physically, the propulsion process is based on two mechanisms which occur 
simultaneously, that is, radiation-pressure \cite{ra,rb} and photophoresis \cite{ashk}.  
In order to navigate  the particle, the propulsion becomes active, that is, the laser is switched on only when the particle 
orientation $\hat{\bf e}(t)$ has the desired direction \cite{chicos2} given by the connection vector between the particle and the target. This clearly requires
monitoring $\hat{\bf e}(t)$ in real time. 

Here we propose an alternative strategy where the
particle orientation is estimated via the difference between the actual and delayed position.
This is similar in spirit to what we have proposed within the optical trapping strategy (see, e.g., eqn.~(\ref{amu}), with the difference 
that we now require {\em two} delayed coordinates instead of only one (due to the absence of a confinement in $y$-direction).
Specifically,
the estimated orientation vector is written as
 \begin{equation}
\textbf{p}(t)   = \left(x(t)-x(t-\delta)\right)\hat{{\bf i}}+ \left(y(t)-y(t-\delta)\right)\hat{{\bf j}},
\label{vectors}
\end{equation}
where $\hat{{\bf i}}$ and  $\hat{{\bf j}}$ are again unit vectors in $x$- and $y$-direction.

\begin{figure}
\scalebox{0.45}{\includegraphics{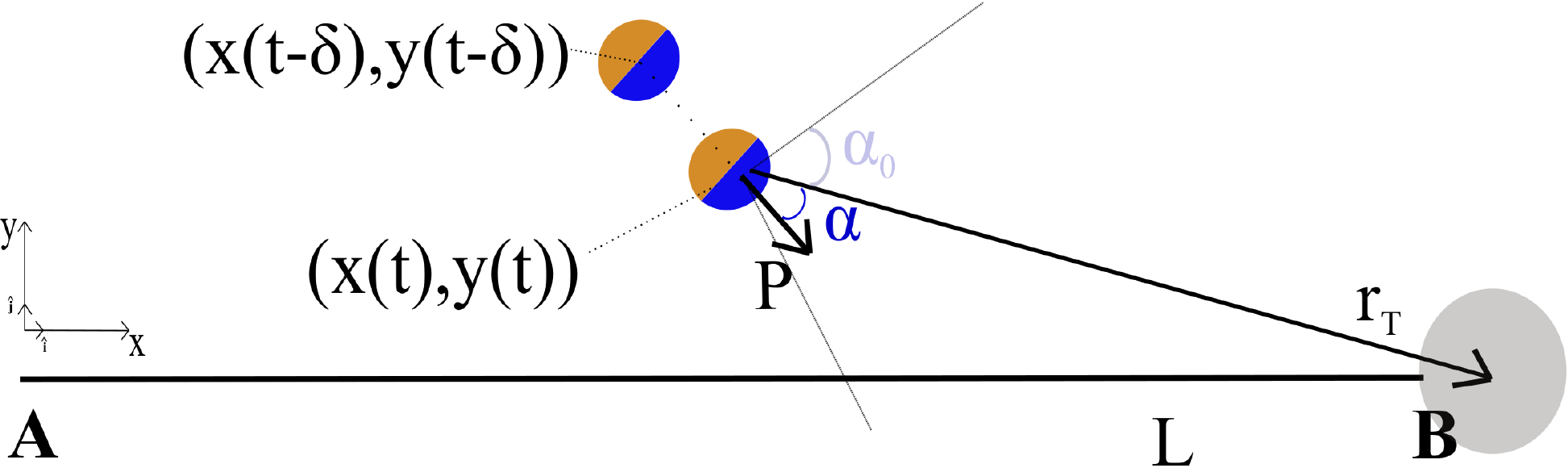}}
\caption{Schematic explanation of the proposed approach in the photon nudging method. 
The intensity of the laser is linearly modified with the angle between the heading vector,
and the 
direct line which connects the particle to its destination.}
\label{etod}
\end{figure}

To quantify the deviation between the particle orientation and the desired direction of motion, 
we introduce the (dimensionless) angle $\alpha(t)$ defined as 
 \begin{eqnarray}
\alpha(t) =  \arccos \frac{\textbf{p}(t)\cdot\textbf{r}_T(t)}{|\textbf{p}(t)| |\textbf{r}_T(t)|},
\label{alpa}
\end{eqnarray}
where the vector $\textbf{r}_T(t)$ points from the actual particle position towards the position of the target (B). Specifically, it is defined as
 \begin{eqnarray}
\textbf{r}_T(t) = (L-x(t)) \hat{{\bf i}} -y(t) \hat{{\bf j}} .
\label{hvectors}
\end{eqnarray}
An illustration of these quantities is given in Fig.~\ref{etod}. The central idea of control is to adapt the laser intensity $I$ based on the actual value of $\alpha(t)$.

Let us now turn to the formulation of the equations of the motion. Similar to our approach for optical trapping (see, in particular, eqns.~(\ref{mom}) and (\ref{mm})),
we assume a linear relationship between the laser intensity (and thus, the motility) and the control parameter, in this case $\alpha$. 
In the optimal case, $\alpha$ should be zero (i.e., $\textbf{p}(t)$ is directed towards the target). 
In order to prevent considerable motion in the direction opposite 
 to the target, we define a cut-off angle $\alpha_0$ above which the intensity (and thus the motility) is reset to a constant small value. 
 With these considerations in mind, we make the following {\em ansatz} for the motility:
 \begin{equation}
v(t) = v_0 \left(1+f(\alpha) \frac{\alpha_0-\alpha(t)}{\alpha_0}\right).
\label{velot}
\end{equation}
In eqn.~(\ref{velot}), the dimensionless function $f(\alpha)$ is set to zero for $\alpha(t)> \alpha_0$ and to a constant positive value $f_0$ when $\alpha(t) < \alpha_0$. 
The resulting motility becomes maximal ($v(t)=v_{max}=v_0(1+f_0)$) if $\alpha(t)=0$, as it should be.

To summarize, we now have two control parameters (contrary to our optical trapping strategy): 
First, the parameter $f_0$ which  determines how fast the particle moves when it has the proper orientation. 
Second,
the cut-off angle $\alpha_0$ which, as we will see from the numerical results presented below, plays a crucial in the control process. 

The Langevin equations of such a controlled motion then read 
\begin{eqnarray}
\dot{x}(t)&=&v(t) \cos\phi(t)+\sqrt{2D_T} \xi_{T,x}(t) \nonumber \\
 \dot{y}(t)&=&v(t) \sin\phi(t) +\sqrt{2D_T} \xi_{T,y}(t) \nonumber \\
\dot{\phi}(t)&=&\sqrt{2D_R} \xi_R(t)
\label{1p}
\end{eqnarray}
with $v(t)$ given by eqn.~(\ref{velot}).
\subsubsection{Simulations}
In the following we present results from numerical simulations of eqns.~(\ref{1p}). Our main aim is to explore to which extent 
the control based on the {\em estimated} orientation $\textbf{p}(t)$ (see eqn.~(\ref{vectors})) can reproduce corresponding results based on
the true orientation $\hat{\bf e}(t)$. The latter is a direct output of our simulations (or "real-time" experiments).

\begin{figure}
\scalebox{0.33}{\includegraphics{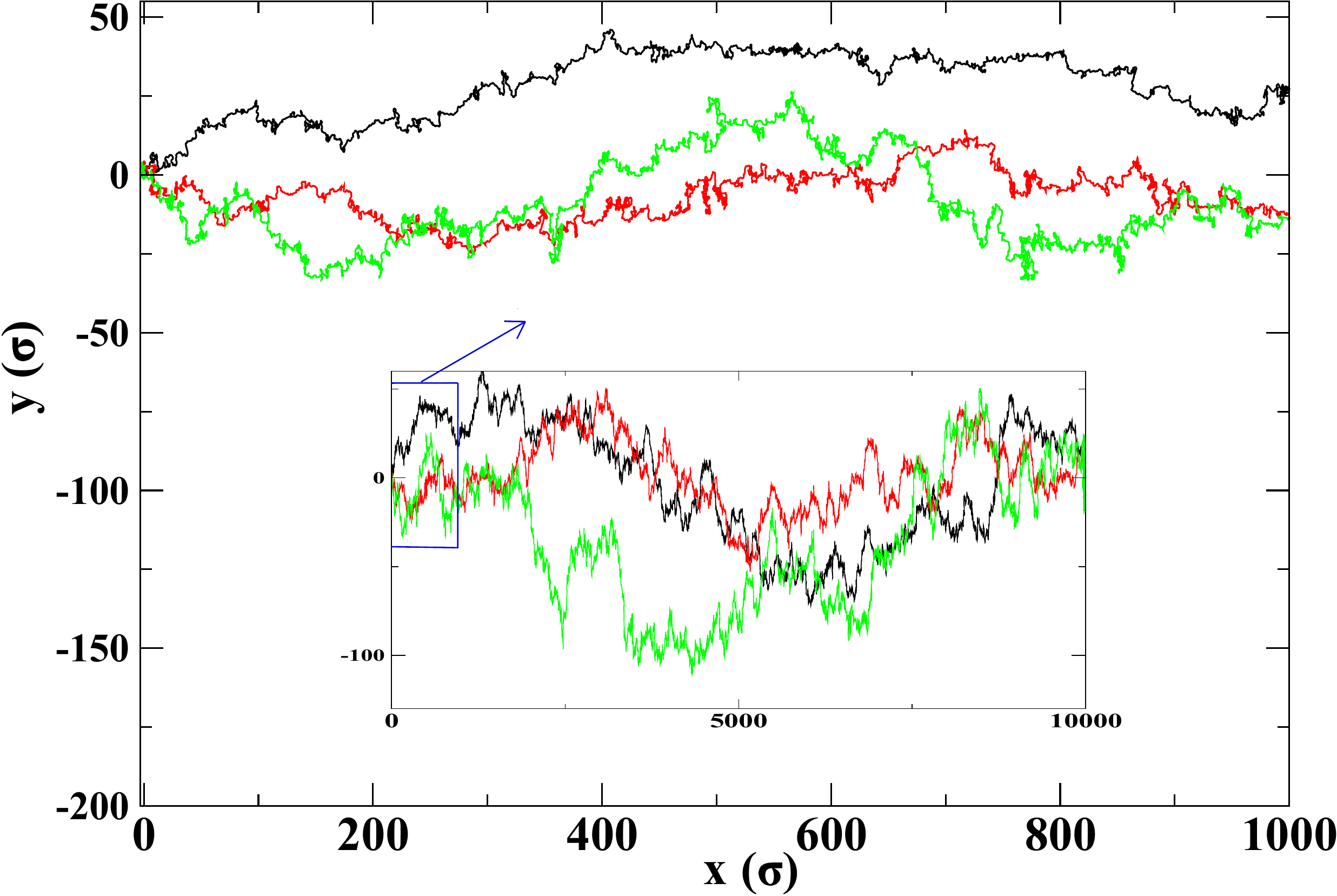}}
\caption{Exemplary particle trajectories in the $x$-$y$-plane for cut-offs
$\alpha_0=15$ (black), $30$ (red), and $45$ (green). Inset: Full trajectories from the starting point to the destination. The remaining parameters are set to $f_0=7$,
$\tau_R= 65\delta$ and  $v=0.1\sigma/\delta$.} 
\label{5et}
\end{figure}

To this end we have performed calculations for different values of the cut-off angle $\alpha_0$ 
at fixed $f_0 = 7$, $v_0= 0.1 \sigma/\delta$ and $\tau_R=65 \delta$. Exemplary trajectories in the $x$-$y$ plane are shown
in Fig.~\ref{5et}. The particle starts at ${\bf r}_A=(0,0)$ and is supposed to move to ${\bf r}_B=(L,0)$.
The data reveal several effects. For small cut-off values (e.g., $\alpha_0=15$) the trajectory involves significant portions in the wrong (i.e., negative $x$-)direction.
This changes upon increase of $\alpha_0$, indicating that higher values of the cut-off parameter provide a faster steering process. 
Finally, the full trajectories presented in the inset of Fig.~\ref{5et} show that the particle reaches its destination for all values
of $\alpha_0$ considered. This indicates that our control based on the estimated particle orientation is indeed successful and robust against changes
of $\alpha_0$.
%
%

\begin{figure}
\scalebox{0.33}{\includegraphics{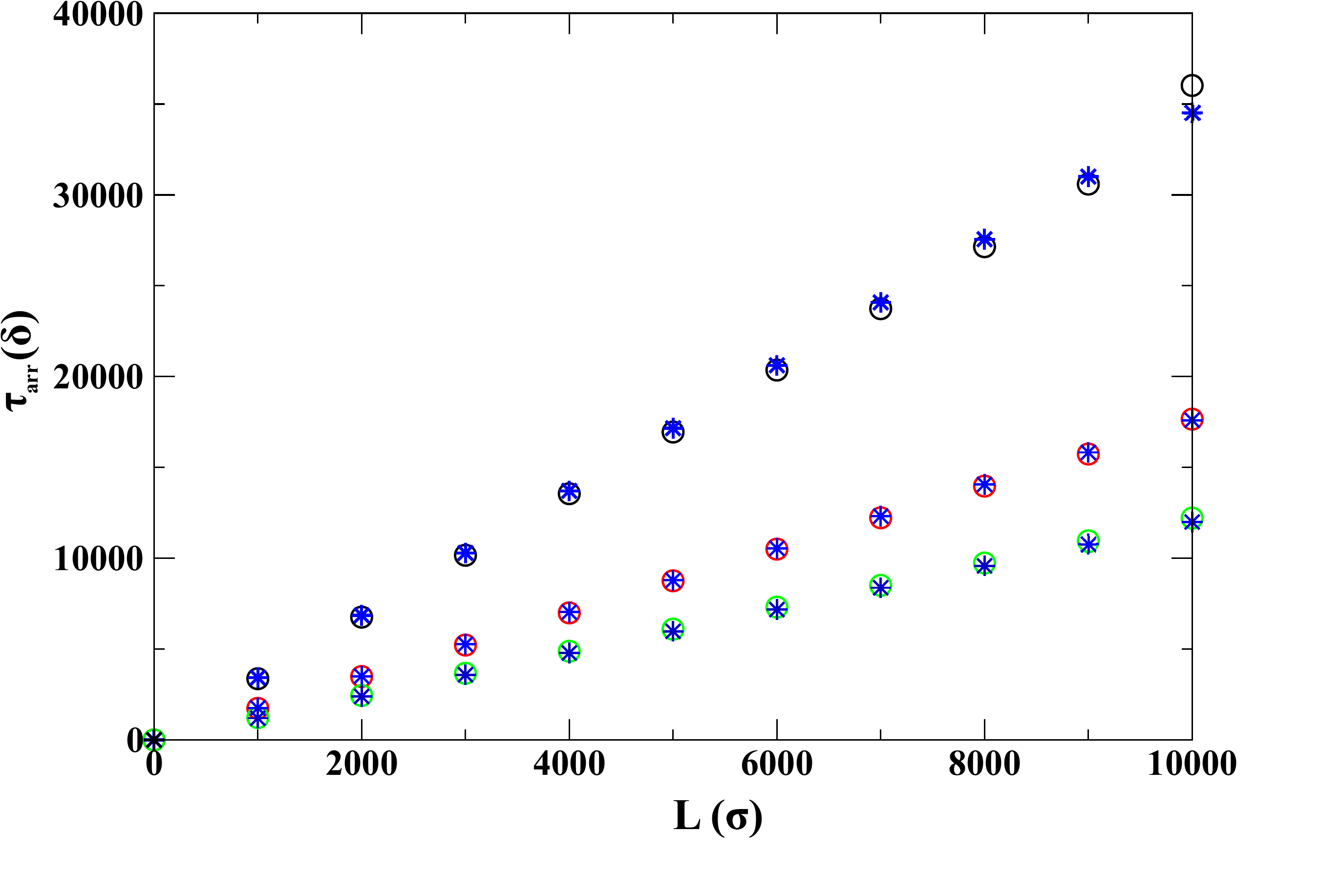}}
\caption{Mean arrival times as function of distance to the target and different values of the cut-off parameter: $\alpha_0=15$ (black), $30$ (red), and $45$ (green).
The data labeled by circles have been
obtained by the present method based on the estimated particle orientation, while those labeled by the blue asterisks
are calculated by the conventional method based on the true orientation. The remaining parameters are set as in Fig.~\ref{5et}. }
\label{6et}
\end{figure}

To compare the method proposed here with the "conventional" strategy based on the true orientation vector $\hat{\bf e}(t)$, we calculated
the mean arrival time. Results for different $\alpha_0$ are presented in Fig.~\ref{6et}.
We find that the present method yields essentially the same results as the conventional method; indeed, deviations
are visible only for the smallest value of $\alpha_0$ (black data). The results also confirm our earlier observation, namely,
that higher values of the cut-off parameter lead to faster steering.

Finally, we mention one conceptual difference between the strategy proposed here and the conventional one: in our case, the 
laser intensity should not never be zero. The underlying reason is that our strategy uses 
the displacements at earlier times to predict the orientation of the particle at a given time (see eqn.~(\ref{vectors})).
Whenever the so-obtained heading vector is oriented in the wrong direction, the particle will still slightly move.
The corresponding  displacement must be detectable (for a camera) in order to allow for corrections at later time. This requires 
a certain minimum motility $v_0$ and thus, intensity $I_0$. On the other hand, $I_0$ should be 
as small as  possible to avoiding significant motion in unwanted 
direction. For small cut-off values $\alpha_0$, this fact could make a non-negligible difference in efficiency compared
to the conventional method, since the particle spends more time in the state 
where the laser would be off in the conventional method. 

\section{Summary}
\label{s5}
In this work, we have explored methods to navigate an active particles through its approximate orientation vector determining the direction of its motion.
The approximation involves the difference between the actual particle position at time $t$ and that a somewhat earlier ("delayed") time, $t-\delta$. 
This approximation is inspired by the idea that, especially for small particles, real-time monitoring of the true orientation can be experimentally very difficult or even impossible. In contrast, positional
control can given achieved via fluorescence spectroscopy even for small particle sizes on the nanoscale. 

We have applied (on a theoretical and numerical level) the idea of using the delayed position for navigation of the particle, first, in the context of optical trapping. By following the particle with a laser trap along the direction
towards the target, we confine its motion effectively into a channel. Navigation in the channel is achieved by introduced an asymmetry in motion based on the approximate orientation. 
The resulting set-up drives the particle efficiently into the desired direction, as we have shown by numerical simulations of the full (delayed) equations of motion
and by analytical theory. The latter is based on a coarse-graining approach for the limit of small delay times, yielding explicit results for the effective force acting on the
particle and the mean arrival time. The agreement between theory and simulation is excellent. In this context, we also note an interesting effect
of the dimension of rotational noise. Indeed, most of our results refer to a completely two-dimensional situation, where the active particle is spatially confined to a plane and rotates only in this plane. 
In section~\ref{sec:spherical}
we have additionally explored the situation that the confined particle can rotationally explore all directions on the unit sphere. It 
turns out that the mean arrival time increases, as one might expect.
Interestingly, this effect can still be captured by a coarse-grained theory as long as the translational motion remains one-dimensional.

As a second application we have considered a variant of the photon nudging method where, instead of the true particle orientation, the approximate one is used. We have provided numerical results
for different values of the cut-off parameter used to adapt the (laser) intensity. The data indicate a very good performance of the approximation. 

We note that, although we have assumed the delay to be small, it is clearly a crucial ingredient: without delay, our approximation
for the particle orientation breaks down. In this sense, our approach provides an example of a feedback-controlled system in which time delay has a constructive effect.
Indeed, in many studies of feedback-control, delay is rather considered as a disturbance, whose role is therefore neglected. Here, not only we do not neglect the delay, 
but also utilize it.

 Of course, it would be very important and interesting to see the performance of our proposal in a real experiment.
In this context, we also mention that there are some ingredients of our proposal which could be applied to a passive particle as well.
In particular, trapping a passive particle by a laser beam is nowadays a standard method (optical tweezer) \cite{ashkin2006optical, dienerowitz2008optical}, and also 
moving traps are quite common \cite{simmons1996quantitative, daly2015optical, gao2017optical}. Furthermore, the approximation of the particle's velocity (which, for our active particle model, equals the orientation vector)
through its displacement vector related to a given time interval between $t$ and $t-\tau$ could also be applied to a passive particle.
One should note, however, that the typical diffusion time scale of a passive particle is smaller than that of an active one,
which might render the approximation more severe. Moreover, for a light-sensitive active particle, changing the intensity of the laser
beam has an impact on the motility, and we have used this fact both, in the optical trapping part and within the photon-nudging part.
For a passive particle, this effect is obviously absent, and one would need another mechanism to drive the particle.

The present work may be considered as a contribution to ongoing efforts to understand and put forward the role of feedback control for stochastic Langevin systems, in this case self-propelled
particles. There are many intriguing open questions, such as thermodynamical implications, which proves to be particularly challenging in presence of time delay \cite{loos2}.
Moreover, from the physical (and applicational) side there is strong interest in navigating the motion not only of 
single self-propelled objects, but also of larger ensemble which can display complex collective behavior
already in the absence of any control. In these contexts, time delay may again play a significant role, as first studies indicate \cite{Mijalkov}.

    
 \section*{Acknowledgment}
We gratefully acknowledge discussion with R. Klages and F. Cichos. This work was funded by the Deutsche Forschungsgemeinschaft (DFG, German Research Foundation) - Projektnummer 163436311 - SFB 910.


\bibliography{rsc-articletemplate} 
 \bibliographystyle{unsrtabbrv}  

\end{document}